\documentclass[11pt]{article} 
\usepackage{epsf}
\usepackage{times} 
\usepackage{graphicx} 
\newcommand{\half}{\frac{1}{2}}
\newcommand{\ie}{{\it i. e. }}
\newcommand{\ea}{{\it et al. }}
\newlength{\pcm}
\newlength{\pmm}
\newcommand {\vertex}{\,\epsfxsize=2.5\pcm \parbox{1.5\pcm}{\epsfbox{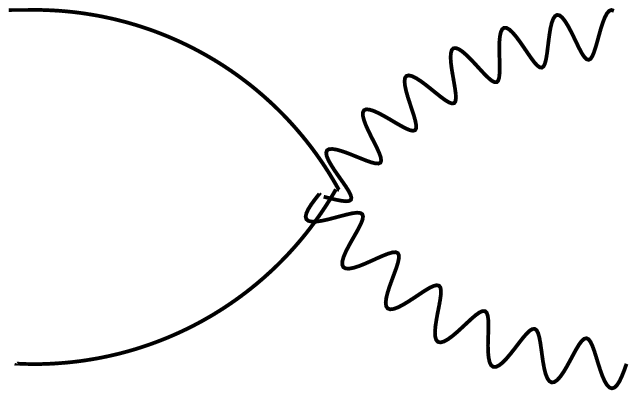}}\,}
\newcommand{\kk}{\hspace{-2.6\pmm}}
\newcommand{\nn}{\nonumber}
\title{Multiplicative Noise in Non-equilibrium Phase Transitions: A tutorial} 
\author{Miguel A. Mu\~noz\\
Instituto de F{\'\i}sica Te\'orica y Computacional Carlos I,\\
Facultad de Ciencias, Univ. de Granada, 18071-Granada, Spain.}
\begin{document} 
\maketitle   

\begin{abstract} 
Stochastic phenomena in which the noise amplitude is proportional to the
fluctuating variable itself,
 usually called {\it multiplicative noise},
 appear ubiquitously in physics, biology,
economy and social sciences. 
The properties of spatially extended systems with 
this type of stochasticity, paying special attention to 
the  {\it non-equilibrium phase transitions} these systems may exhibit,
 are reviewed here. 
In particular we study and classify the possible universality classes of 
such transitions, and discuss some specific physical realizations including
depinning transitions of non-equilibrium interfaces, 
non-equilibrium wetting phenomena,  synchronization of spatially extended 
systems and many others. 
\end{abstract} 

\section{Introduction: Multiplicative Noise}
 Since the pioneering description of Brownian motion as a stochastic process,
Langevin representations of statistical systems have played a central role in the 
development and understanding of statistical mechanics \cite{VK,Gardiner,HH}.
The introduction of noise on otherwise deterministic (noiseless) systems have proven 
to be essential in order to properly describe numberless problems in physics, 
chemistry, biology, economy or social sciences. 
In some cases noise may lead to rather surprising and counterintuitive behaviors. 
A well known example of this is {\it stochastic resonance} in which the response of a 
stochastic system to an external periodically oscillating field is enhanced by increasing 
its internal noise amplitude \cite{SR}.
Other instances are resonant activation \cite{SA},
noise-induced spatial patterns \cite{Sancho,n1}, and noise induced
ordering transitions (NIOT) \cite{HL,NIOT} to name but a few.

Most of the first tackled examples, including Brownian motion,
were properly described by Langevin equations with {\it additive noise}, 
\ie state-independent fluctuation amplitudes.
However, in order to properly account for more complex situations {\it multiplicative noise} 
(MN),
\ie a variable-dependent noise amplitude, might be (and, actually, is) required. 
A prototypical example of this are stock market fluctuations in which the
dayly variation of a given company quotation is given by a certain {\it percentage} of 
its initial value, \ie variations are larger for larger starting  prices.
 Another example are time fluctuations in the number 
of molecules of a given species present in a catalytic reaction: the fluctuation amplitude 
depends obviously on the number of molecules at each time. In the same way, 
fluctuations in population dynamics models as the Lokta-Volterra can be modeled
by multiplicative noise \cite{PS}.
Also, situations in which some control parameter fluctuates may generically
lead to the presence of MN \cite{Sancho,Vi}. Some more example are the equations
describing the  fluctuations of advancing fronts \cite{Tripathy}, or Schr\"odinger 
equations in the presence of disordered potentials \cite{Kani}.

It is well established by now that many of the situations in which noise
generates counter-intuitive features can be modeled by Langevin equations
with MN.
 As a rule of thumb, in systems with additive noise, the smaller the noise amplitude, 
the more ``ordered'' the stationary state. On the other hand, 
for systems with MN this is not necessarily the case, 
and, as said before, a wealth of instances have been described where the 
degree of ordering is enhanced upon enlarging noise amplitude; 
noise induced phase transitions are a good example 
\cite{HL,Sancho,NIOT,BK,ReTr,GMS}.

Before proceeding further, 
let us stress that among many other applications, Langevin equations 
(with or without MN)
have turned out to be a highly convenient theoretical tool for the analysis and 
characterization of critical phenomena, including both static and dynamical aspects; an
application that will be of special interest along this paper.
For instance, models A and B
(both of them defined by Langevin equations \cite{HH}) have become paradigmatical 
examples describing the universal features of Ising-like phase transitions,
with non-conserved and conserved dynamics respectively, in the simplest possible way
\footnote{Other symmetries and/or conservation laws can be easily implemented in
this language, generating a zoo o Langevin models: C, D,... H. \cite{HH}}.

  In this review we focus on phase transitions in spatially extended systems 
with multiplicative noise. We will explore their phenomenology, paying special 
attention to their critical properties, and will review some applications to 
physical systems, including pinning-depinning, wetting problems 
and synchronization transitions, among many others.

\section{Multiplicative Noise basic equation}

 In has been recently shown \cite{GMS} that the following
multiplicative noise (MN) Langevin equation intended in the {\bf Stratonovich} 
sense \footnote{See Appendix A for a brief discussion of the Ito versus 
Stratonovich dilemma.} is the minimal model for noise-induced ordering transitions 
in spatially extended systems \cite{NIOT,BK}:
\begin{equation}
{\partial}_{t} \phi(x,t) = - a \phi(x,t) - b \phi^{p} (x,t) + 
D {\nabla}^{2} \phi(x,t) +  \sigma  \phi(x,t) \eta(x,t).
\label{MN}
\end{equation}
$\eta(x,t)$ is a Gaussian white noise, $\phi(x,t)$ the order-parameter 
field that usually represents a density  or an activity-degree field,
  $a, b > 0, D > 0$ and $\sigma > 0$ are parameters, and $p$ is an integer constant obeying
$p \ge 2$ (typically $p=2$ or $3$). As discussed in \cite{GMS}, for this equation
the larger the noise amplitude the more ordered the stationary state.
A very similar equation has been proposed by Pikovsky and Kurths as a simple stochastic
description of synchronization transitions (ST) in extended systems \cite{PK,synchro}
\footnote{To  the best of our knowledge this is the first time that Eq.(\ref{MN})
appeared in the literature.}
(see also section \ref{App}). 

Eq.(\ref{MN}) exhibits, among other non-trivial features, a non-equilibrium phase transition
as the control parameter $a$ is varied, that has been profusely studied 
\cite{MN1,MN2,MN3,MN4}.
In particular, for $a>a_c$, where $a_c$ corresponds to a transition point, the 
only stationary state is $\phi(x)=0$, defining  an {\it absorbing phase}
 \footnote{Sometimes we will refer to this as
``quasi-absorbing'' phase for reasons that hopefully will become clear in what follows.}.
On the other hand for $a<a_c$ the system seats in an active phase with 
$\langle \phi \rangle \neq 0$. This defines an absorbing-state type of 
phase transition. 

Even if, strictly speaking, a MN Langevin equation is any in 
which the noise is a certain function of the stochastic variable, in what follows, 
we restrict the adjective {\it multiplicative} to describe a noise linear in the field,
as the one in Eq.(\ref{MN}).
This excludes explicitely, among many other cases, the Reggeon field theory 
Langevin equation in which the noise amplitude is proportional to the 
square-root of the variable (see below).

Let us start the analysis of Eq.(\ref{MN}), 
the most basic MN equation, \footnote{The effect
of MN has been studied also in other type of Langevin equations, as
for example the Swift-Hohenberg one \cite{n1,Cross}, 
that has become in paradigm in the study of
pattern formation, or equations with conserved order-parameter as the Cahn Hilliard one
\cite{Cross,Sancho}. We will not explore these cases in the present review.}
by describing its three contributions: 
\begin{enumerate}
\item
 A single-site deterministic force, $F (x,t)= - a \phi(x,t) -  b \phi^{p}(x,t)$,
 which can be derived from the potential
 $V[\phi]  = {a\over 2} \phi^{2}  + {b \over p+1} \phi^{p+1}$. 
This is the simplest potential describing a continuous bifurcation 
(transition) from a $\phi=0$ state (for $a>0$) to a non trivial state, 
$\phi \neq 0$ (for $a<0$)  
\footnote{This is usually called a {\it transcritical bifurcation} in the 
literature of dynamical systems.}.
Higher order terms could be included in $V(h)$ but, as it is well known in the 
folklore of critical phenomena, they are expected to be irrelevant 
in what respects universal properties, 
at least as long as $b$ remains positive, ensuring stability.

If for symmetry reasons the order parameter of the problem under consideration 
exhibits a $Z_2$ inversion (or plus/minus or up/down) 
symmetry then an odd value of $p$ (typically $p=3$) is required,
\footnote{Observe that if $p=3$ the order parameter can be
either positive definite, or defined in the whole real axis. This should not
affect critical properties.}
while $p=2$ will be used by default in the absence of such a symmetry as the most 
basic choice.

\item
  A diffusion term, $ D {\nabla}^{2} \phi(x,t)$, which, as usual,
is the force derived from a surface tension 
$ D  \int d^d x \half (\nabla \phi)^2$ tending
to smooth away fluctuations.
Higher order terms in a gradient expansion
could also be introduced without altering the critical properties.

\item
  A noise term proportional to $\phi(x,t)$, \ie a MN.
\end{enumerate}

In order to put the problem under perspective, and for pedagogical reasons, 
let us compare this Langevin equation to other very well-known ones:

{\bf(A)} Model A describing Ising-like phase transitions with non-conserved dynamics
\cite{HH,ZJ,Cardy}:
\begin{equation}
   {\partial}_{t} \phi(x,t) = - a \phi -
   b \phi^3 + D {\nabla}^{2} \phi(x,t) +  \sigma \eta(x,t).
\label{ModelA}
\end{equation}
Apart from the cubic term imposing up-down symmetry, the only difference
with Eq.(\ref{MN}) is in the noise amplitude, constant in this case.  

{ \bf(B)} The Reggeon-Field theory (RFT) known to describe phase transitions 
in a vast class of systems with absorbing states
(directed percolation (DP)  and the contact process being
paradigmatic instances \cite{RFT,Kin,Haye,MD,Granada,Odor}),
\begin{equation}
   {\partial}_{t} \phi(x,t) = - a \phi -
   b \phi^2 + D {\nabla}^{2} \phi(x,t) +  \sigma \sqrt{\phi(x,t)} \eta(x,t).
\label{RFT}
\end{equation}
The only difference with Eq.(\ref{MN}) is in the noise amplitude, proportional here
to the square-root of the activity field. The difference between the absorbing states
in this class and the ``quasi-absorbing'' states of the MN family 
is discussed in detail in \cite{nature} (see also Appendix B).
 In a nutshell the difference is that 
``true'' (RFT) absorbing states correspond to a non-integrable singularity at the 
origin ($\phi=0$) of the single-site probability distribution function
 (present  even in the active phase)
\footnote{This means that the only true stationary solution is a Dirac-delta 
distribution at $\phi=0$.}
 while such a singularity does not exist or is integrable
in the active phase of problems in the MN family.

{\it The three equations presented so far, with noise amplitudes $\phi^0$, $\phi^{1/2}$ and 
$\phi^1$ respectively,
 represent three main groups of phase transitions occurring in systems 
without anisotropies, long-range interactions, 
extra components, non-Markovian effects, or quenched disorder}.  
An ever-growing number of critical phenomena in very different contexts 
have been reported to belong to any of these very robust three classes. 
  While for both the Ising class (Model A), and the directed percolation (RFT) classes
several comprehensive reviews exist in the literature (see for example 
\cite{HH,ZJ,Haye,MD,Odor}) to the best of our knowledge there is no compilation of
 known results for the MN equation and variations of it.
It is the purpose of this review to report on the interesting and rich phenomenology
of this and closely related systems, as well as on their
many physical realizations.  

\section{Connections with related problems}

Before analyzing the properties of Eq.(\ref{MN}), for
the sake of generality, we enumerate now its connections with other
relevant problems in non-equilibrium statistical physics.
In particular it is related to:

\vspace{0.2cm}
   {\bf 1. The Kardar-Parisi-Zhang (KPZ) equation} 

The KPZ equation describes generically the roughening of 
interfaces under non-equilibrium conditions and has attracted an 
enormous deal of attention in the last decade \cite{HZ,Laszlo,Krug}. 
To see how the connection between MN and KPZ works, 
 Its connection with the MN equation comes out by performing
the so-called Cole-Hopf transformation 
$\phi(x,t)=\exp[h(x,t)]$ to Eq. (\ref{MN}):
\begin{equation}
\partial_t h(x,t) = -a - b ~e^{(p-1)~h} + D \nabla^2 h + D (\nabla h)^2 
+ \sigma \eta(x,t)
\label{KPZ}
\end{equation}
which is the celebrated KPZ equation, describing an interface of height $h(x,t)$
(with respect to a reference substrate at $h=0$)
in the presence of an additional bounding term,
$- b ~e^{[(p-1) h]}$, pushing regions where $h(x,t)>0$ towards negative values. 
The first two terms can be written as derived from a potential 
$V(h)= a h + {b\over (p-1)} e^{[(p-1) h]}$ (observe that $a$ controls the 
potential slope for large values of $|h|$).
Eq.(\ref{KPZ}) is usually called in the literature a KPZ with {\it an upper wall}.
If instead we had defined $\phi=\exp(-h)$, the sign of the emerging  
KPZ-non-linearity would be negative, inducing a {\it lower wall}
(\ie the argument of the exponential bounding term changes sign, and therefore 
negative values of the height variable are pushed towards positive ones).
  In other words, 
{\it a KPZ interface with positive non-linearity in the presence of an upper wall is 
completely equivalent to a KPZ with negative non-linearity and a lower wall}. 
These two cases can be mapped one into the other and are described by Eq.(\ref{MN}). 
We will refer to this as Multiplicative Noise 1 ({\bf MN1}) class. 

The active phase, $\langle \phi \rangle \neq 0$, is translated in the
interface language into a pinned-to-the-wall phase 
while the absorbing
one maps into a depinned phase (with interfaces escaping towards infinity or
minus infinity at a constant average velocity). 
The absorbing-active phase transition becomes a 
pinning-depinning one within this language.

 The curious reader could pose himself/herself the following question: 
what happens if we consider a KPZ interface in the two remaining cases, \ie
 with positive non-linearity in the presence 
of a lower wall, or with a negative non-linearity and an upper wall? 
Can these problems be described in terms of a MN equation? 
Taking the following  KPZ-like equation with an upper wall ($~  b ~e^{(p-1) h}$)
and a negative KPZ-nonlinearity,
\begin{equation}
\partial_t h(x,t) = - a - b ~e^{(p-1)h} + D \nabla^2 h - D (\nabla h)^2 + \sigma
\eta(x,t)
\label{KPZ2}
\end{equation}
as starting point, and
performing the inverse Cole-Hopf transformation we arrive at:
\begin{equation}
{\partial}_{t} \phi(x,t) = - a \phi -
b \phi^p + D {\nabla}^{2} \phi(x,t) - 2 D {(\nabla \phi(x,t))^2 \over  \phi(x,t)} + 
\sigma  \phi(x,t) \eta(x,t)
\label{MN2}
\end{equation}
which is identical to Eq.(\ref{MN}) except for an extra singular term.
 Eq.(\ref{MN2}) is a {\it modified} MN equation, that we will name
 Multiplicative Noise 2 ({\bf MN2}) equation. 
Obviously the same result is obtained for a lower wall and a positive
KPZ non-linearity by using $\phi=\exp(-h)$.
As we will illustrate in the forthcoming sections the extra term 
$- 2 D {(\nabla \phi(x,t))^2 \over  \phi(x,t)} $, which 
can also be written as 
\begin{equation}
- 2 D \large(\nabla \phi(x,t)\large).\large(\nabla \ln(\phi(x,t))\large) = \pm 2 D
  \large(\nabla \phi(x,t)\large) \cdot \large(\nabla h(x,t)\large)
\label{log}
\end{equation}
 (apart from being singular) introduces a relevant perturbation making 
MN2 essentially different from MN1 \footnote{Observe that the presence of
$\nabla h(x,t)$ in the equation for $\phi$'s Eq.(\ref{log}) 
means that MN2 finds a more
natural representation in terms on $h$'s than in terms of $\phi$'s.}.

{\bf Summing up}: for a fixed value of the KPZ non-linearity the introduction of limiting
upper or lower walls lead to essentially different physics. For a fixed wall, changing  
the sign of the non-linearity also changes essentially the behavior.

\begin{figure}
\begin{center}
\includegraphics[width=0.6\textwidth,height=0.3\textheight]{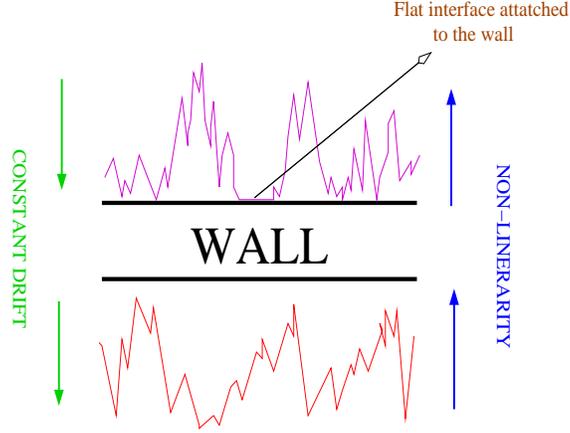}
\caption{Schematic representation of KPZ interfaces with a positive non-linearity
in the presence of upper  ({\bf MN1})  and lower ({\bf MN2}) walls respectively. 
The walls are depicted as rigid lines (corresponding to $p \rightarrow \infty$; for 
finite values of $p$ the walls become exponential cutoffs).
Observe the difference between the two cases: while in one of them the constant driving 
force 
pulls the interface away from the bounding wall (MN1), in the other,
the interface is pushed to the wall (MN2).}
\label{walls}
\end{center}
\end{figure}
 Let us illustrate the physical reasons behind this fact. For that, we fix
the sign of the non-linearity to be positive. For an upper (lower) wall the non-linearity 
pulls (pushes) the  interface from (against) the wall, while the constant term, 
acts in the opposite direction in order to keep the interface pinned. 
These are the two only mechanisms driving the interface average velocity, which 
imposing stationarity is given by 
 \begin{equation}
\langle \partial_t h  \rangle =~ 0 ~ =  
\pm a \pm \langle b ~e^{\mp h} \rangle + \langle D (\nabla h)^2 \rangle. 
\label{KPZstat}
\end{equation}
At the critical point $\langle e^{\mp h} \rangle =0$ and, therefore,
$a$ has to compensate $\langle  D (\nabla h)^2 \rangle$. 
Let us now consider a flat interface ($(\nabla h)^2=0$)
at the critical point. While in one case
the constant drift, $a$,  pulls it from the wall, causing an effective repulsion from it,
in the other one the flat piece is pushed against the wall, being therefore stable
(see figure \ref{walls}).
This mechanism leads to essentially different stationary probability distributions
in both cases, and ultimately to different universality classes \cite{MN3}.

The limit $p \rightarrow \infty$ describes
a {\it hard} (impenetrable) wall: the interface
cannot cross the $h=0$ limit in either the upper ($h(x,t) < 0$)
nor the lower wall case ($h(x,t) \geq 0$).  As we will show 
the ``hardness'' (or rigidity) of the wall is not a relevant feature. 
Indeed, $p$ can be reabsorbed from the KPZ-like equation by modifying 
the Cole-Hopf transformation to a more general form: $\phi=\exp(m h)$, 
where $m$ is a conveniently chosen constant.

\vspace{0.2cm}
 {\bf 2. Directed polymers in random media (DPRM)} \cite{HZ}.

The connection with directed polymers in random media has also been known for some time.
Indeed, the partition function of the DPRM can be easily seen to obey the
following equation (imaginary time Schr\"odinger equation) \cite{HZ}
\begin{equation} 
\partial_t Z (x,t) = [  D \nabla^2 + \eta(x,t) ]  Z (x,t)
\label{Scho}
\end{equation}
including both: an elastic term (diffusion) and a random potential. 
Therefore any problem in which the possible values of the DPRM free energy, 
$F \propto  \ln Z$ are restricted, owing to some unspecified reason,
from above or from below, can be mapped into a MN equation, 
or into the KPZ equation with an upper or lower wall respectively. 
In particular, it has been pointed out by Hwa and coauthors 
that algorithms used routinely 
in molecular biology for sequence local-alignment are straightforward realizations
of DPRM with a free-energy constraint and are related to 
the KPZ-problem with a lower wall \cite{Hwa,MN3}.  
Other problems in {\it condensed matter physics}, in which a Schr\"odinger equation 
appear including a random potential, can be ascribed to this class; this can
include problems related to Kerr dielectric guides, propagation of
optical pulses, multi-channel optical fiber networks, etc. 
See \cite{Kani} and references therein. 

\vspace{0.2cm}
  {\bf  3. Noisy Burgers equation of turbulence}  \cite{Burgers,HZ}.

 Considering  a KPZ equation and transforming $v(x,t)=\nabla h (x,t)$ one obtains
the (noisy) Burgers equation describing a randomly stirred  vorticity-free
(\ie $\nabla  \times v =0$) fluid  
\begin{equation}
{d v \over dt} \equiv \partial_t v + v \cdot \nabla v = 2 \nabla^2 v + \nabla \eta
\label{Burger}
\end{equation}
where we have fixed $D=2$ to ensure invariance of the total derivative under
rescaling \cite{HZ,Laszlo}. The presence of an extra upper wall
generates an additional term
which cancels out high velocities. 
This velocity constraint can mimic the role of viscosity in some problems in 
fluid-dynamics.

\section{Properties of Multiplicative Noise. The MN1 class}

In this section we explore the phenomenology associated with MN1 by employing
different levels of approximation and tools.

\subsection{Zero dimensions} 
An exact solution of the one-variable (zero-dimensional) case was provided
by Graham and Schenzle \cite{GS} (see also \cite{HL} and \cite{Granada}
 for a review).
Indeed, the stationary 
solution of the associated Fokker Planck equation can be easily worked-out
(see Appendix B)
\begin{equation} 
P_{st}(\phi)= {1 \over \phi^{1+2a/\sigma^2}} \exp{
\left( {-2 b  \phi^{p-1}    \over \sigma^2 (p-1)} \right) }.
\label{MN0d}
\end{equation}
Let us just underline two curious non-trivial properties of this solution \cite{GS}:
\begin{itemize} 
\item All the moments scale in the same way(!) 
  For example, for $p=2$ it is just a matter of algebra to obtain
\begin{equation}
\langle \phi^m \rangle = 2^{m/2} \Gamma \left[ {m + a/\sigma^2 \over 2}\right]
 \Gamma \left[ -a/2\sigma^2 \right],
\label{moments}
\end{equation}
and, therefore, $\beta_m$ defined by $\phi^m \sim |a - a_c|^{\beta_m}$ 
is equal to unity ($1/(p-1)$ for an arbitrary $p$) for all $m$ \cite{GS,MN1}. 
This is a rather extreme case of {\it anomalous scaling}, 
having its roots in the fact that the asymptotic behavior is
controlled by large rare events \cite{Redner,Sornette}.  
\item The susceptibility (response to an external field) diverges not 
only at the critical point, but
in a finite interval in the absorbing phase, and the associated critical
exponent varies continuously \cite{MN1}. 
\end{itemize}

\subsection{Mean Field approaches}
  Developing a consistent and well defined mean-field approximation of 
Eq.(\ref{MN}) is not a trivial task \cite{MN4}. 
It has been only recently that Birner \ea (completing some previous 
partial results \cite{MN4}) have shown that \cite{Birner,Marsili}
\begin{equation}
\beta_1  = \sup \left( {1 \over p-1}, { \sigma^2 \over 2 D}\right).
\label{mf}
\end{equation}
This result can be obtained by approximating the Laplacian in the Langevin 
equation by its mean field analog,
 ${\frac{1}{N}} \sum_{\forall i} (\phi(i,t) - \phi(x,t))
 = \langle \phi \rangle ~-~\phi(x,t)$,
writing the equivalent Stratonovich (see Appendix A) Fokker-Planck equation
\begin{equation}
\partial_t P(\phi,t)=  \partial_\phi \left( [ a \phi + b \phi^p - D (\langle \phi \rangle
 -\phi)] P(\phi, t) \right)
+ {\sigma^2 \over 2 } \partial_\phi~ \left( \phi \partial_\phi~ [ \phi P(\phi, t)]  \right)
\label{FP}
\end{equation}
and then solving it self-consistently in the stationary limit
(see \cite{Birner} and \cite{NIOT,MN4,Marsili}).
Let us underline the non-trivial nature of the mean-field result Eq.(\ref{mf}):
for high noise-intensities it does not coincide with 
the zero-dimensional $\beta$ (previously called $\beta_1$) exponent 
$\beta=1/(p-1)$, as usually occurs. Therefore, coupling to the neighbors is a key
ingredient to build up a sound mean-field description.
At a field theoretical level (see next subsection) no
systematic naive-power counting analysis or tree level approximation
has succeeded so far in reproducing the non-trivial result of Eq.(\ref{mf})
(see \cite{MN4}).
 Other critical exponents, can be easily computed as usual in mean-field
theory. They  are defined through $ \tau \sim \Delta^ { \nu_\perp}$,
     $ \xi \sim \Delta^ { \nu_\parallel}$,
     $ \langle \phi \rangle \sim t^{-\theta}$, 
 or $\tau  \sim \xi^z$, with $\tau$ ($\xi$) being  the correlation time
(length), and $\Delta= |a-a_c|$ the distance to the critical point.  In the current 
 mean field approximation:  
$\nu_\perp=1$, $\nu_\parallel=1/2$, $\theta= \beta  = \sup \{ 1/(p-1),  
\sigma^2/(2 D)\}$, and $z=2$.
 
\subsection{Beyond Mean Field - Perturbative expansion}
In order to properly incorporate fluctuations and go beyond mean field, as usual,
one has to resort to perturbative expansions combined with renormalization group (RG) 
techniques \cite{ZJ,GF}.
 This can be done either directly  at the Langevin-equation level, or alternatively
working with the corresponding generating functional.
Here, contrarily  to \cite{MN1} we choose to employ the second of these frameworks. 
The generating functional associated with Eq.(\ref{MN}) 
\footnote{Here we use the Ito representation, as for it the Jacobian of the employed 
 transformation and the theta function evaluated at $0$ can be seen to be equal
 to $1$ and $0$ respectively, simplifying the algebra \cite{GF}. The calculation 
 would be almost identical in the Stratonovich representation, leading to 
essentially the same 
results, except for a shift in the critical point.} 
is \cite{ZJ}: 
\begin{eqnarray}
&Z[J,H] =  \cal{D} \phi \cal{D}  \psi  \exp(- S[\phi, \psi, J,H])    \nonumber \\
&S = - \int d^d x \int dt  [ {\sigma^2 \over 2} \phi^2 \psi^2 -
\psi[\partial_t \phi + a \phi + b \phi^p - D \nabla^2 \phi] - J \phi - H \phi]
\label{GF}
\end{eqnarray}
where $\psi$ is a response field, $J$ and $H$ sources coupled to
the order parameter field, $\phi(x,t)$, and to the response field, $\psi(x,t)$,
 respectively
\footnote{This functional can be easily worked out, by writing the probability of a given
time-sequence realization of the noise as a product of Gaussian noise distributions
at each time, generating in this way a path integral. Then variables have to be changed 
from $\eta(x,t)$ to $\phi(x,t)$, 
and finally  a dummy (ghost) variable $\psi$ is introduced by means
of a Gaussian transformation, in order to avoid singularities typically induced by the MN.  
Afterwards the sources $J$ and $H$ are introduced by hand. The process is rather
standard; a detailed description can be found in \cite{ZJ} or \cite{GF}.}.  
Naive power counting shows that the upper critical dimension, below which mean
field results cease to be valid, is $d_c=2$ \footnote{This result is easily obtained by imposing 
the noise term coefficient to be dimensionless at $d_c$.}.
Following standard perturbative techniques, one can perform an 
$\epsilon$-expansion below $d_c$.
 The noise amplitude is written in dimensionless form as 
$\sigma^2 \equiv u \mu^\epsilon$, where $\mu$ is an 
arbitrary renormalization scale (for example $\mu^2=a$).
In momentum-frequency space the propagator is ${ (-i \omega + a  +  \kappa^2)}^{-1}$ 
(for simplicity  we fix here $D=1$) and the noise vertex is
\begin{eqnarray}
 \sigma^2\vertex   \nn
\end{eqnarray}
The vertex ($\psi  \phi^p$) is like the previous one but  
with only one wavy line and $p$ straight ones.
 It is easy to see that 
there are no perturbative corrections to the propagator, and
the only possible diagrammatic corrections to the noise vertex have the 
following structure 
\begin{eqnarray}
\sigma_R^2 &= &\sigma^2  \nn \\
       &+& \sigma^4 \vertex\kk\vertex \nn\\
       &+& \sigma^6 \vertex\kk\vertex\kk\vertex \nn\\
       &+& \sigma^8 \vertex\kk\vertex\kk\vertex\kk\vertex
        + \ldots \nn  \\
       &=& {\sigma^2 \over 1 - \sigma^2  \vertex\kk\vertex} \nn
\label{series}
\end{eqnarray}
which can be written as
$\sigma_R^2  = {\sigma^2  \over 1 - \sigma^2 I}$
where I denotes the one loop diagram at zero external frequency and momenta,
and $\mu^2=a$
\begin{equation}
I = {1 \over (2 \pi)^{(d+1)}} \int d^d \kappa~ d \omega~  
 {1 \over i \omega + \kappa^2 
+ \mu^2} ~  {1 \over -i \omega + \kappa^2 + \mu^2} =
 \alpha {\mu^{-\epsilon} \over \epsilon}
\end{equation}
with $\alpha=2/(8 \pi)^{d/2} \Gamma(2-d/2)$ \cite{ZJ,GF,RG-KPZ}. 
The pole $1/\epsilon$ captures the presence of an infrared singularity below $d=2$.
For $d > 2$, $I$ is finite ({\ie } it is infrared convergent) 
\footnote{We assume the existence of a lattice cut-off ensuring the integrals to be 
ultraviolet convergent.}
therefore if $\sigma^2$ is small enough $ \sigma^2 I < 1$ and the mean-field theory 
predictions remain valid perturbatively. 
At $\sigma^2 = I$ a singularity appears in $\sigma_R^2$;
above this noise value the perturbative expansion ceases to be meaningful;
usually it is assumed that values of $\sigma^2  > I$ flow to a non-perturbative 
fixed point, controlling the long-wavelength large-time asymptotic behavior.
For the corrections to vertex $\psi  \phi^p$, one has exactly the same structure, 
except for the 
rightmost noise vertex (see Eq.(\ref{series})) that is replaced by the other one, 
leading to
 \begin{equation}
b_R  =  b  (1 +   \sigma^2 I + \sigma^4 I^ 2  + ...)=   {b  \over 1 - \sigma^2 I} 
=  b \sigma_R^2 /\sigma^2
\end{equation}
Therefore, renormalizing $\sigma^2$ keeps also $b_R$ finite, and no extra
 renormalization is needed. 
As the degree of the non-linearity $p$ does not introduce different 
divergences in the perturbative expansion, it should not affect the critical behavior 
(as long as $p \ge 2$).

 Proceeding with the renormalization program, we compute the Wilson $\beta-$ function
defined as 
\begin{equation}
\beta (u_R) = \mu \partial_\mu |_{\sigma} u_R = -\epsilon u_R - \alpha u_R^2
\label{beta}
\end{equation}
with a unique physical solution, $u_R=0$,
stable for $\epsilon < 0$ (\ie  above two dimensions) with associated mean
field critical exponents.
The renormalization-group flow-diagram is very alike the well-known one for KPZ 
\cite{KPZ,RG-KPZ}, in particular, for $d < 2$ the 
flow runs towards infinity for any noise amplitude, 
while for $d \ge 2$ there is a separatrix between the
trivial fixed-point basin of attraction and the runaway-trajectories, {\ie}
between the weak- and the strong-coupling fixed points.  
\begin{figure}
\begin{center}
\includegraphics[width=0.5\textwidth,height=0.2\textheight,angle=0]{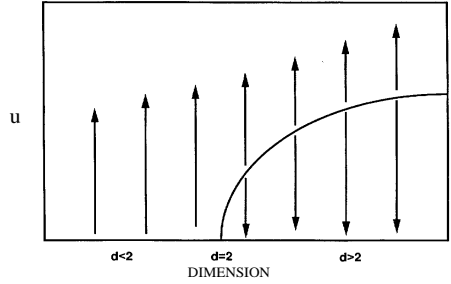}
\caption{Schematic representation of the RG flow of the dimensionless
variable $u$ as a function of dimensionality, as derived from Eq.(\ref{MN}).}
\label{flow}
\end{center}
\end{figure}
Let us emphasize both, the absence of any stable non-trivial perturbative fixed point, 
and the fact that the theory is {\bf super-renormalizable}, in the sense 
that perturbative diagrams
 have been computed and re-summed to all orders.
An alternative derivation of these results can be found in \cite{MN1}.  
See also
\cite{RG-KPZ} for a very similar calculation presented in a pedagogical way.
\footnote{It is illuminating to compare the generating functional 
Eq.(\ref{GF}) with that associated with
the reaction $A + A \rightarrow 0$ (kink-antikink  annihilation) 
\cite{Peliti,Lee,nature,Fock}. It has the same structure as (\ref{GF}) but for
the sign of the noise term. In terms of the  Langevin equation the sign alteration
tantamounts to the introduction of a purely {\it imaginary noise} \cite{nature,Gardiner}.
This simple change has dramatic effects, in particular the Wilson $\beta$-function is as 
Eq.(\ref{beta}) but with a positive sign for the quadratic term,
generating a non-trivial perturbative fixed point below $d=2$ \cite{Peliti}, and a
different non-trivial physics.}

 So far, we have not proven that the strong coupling fixed points in MN
(KPZ plus a wall) and KPZ coincide. \footnote{It is well known that the KPZ
equation should have a strong coupling fixed point controlling its
non-trivial physics \cite{RG-KPZ,Laszlo}.} 
Observe, however, that right at the critical point the interface falls
continuously to minus infinity, as far from the wall as wished.
It is therefore very reasonable to assume, that asymptotically, the dynamical 
exponent should be controlled by KPZ scaling (in the absence of the wall).
In particular, 
it can be shown \cite{MN2} that:  
{\bf (i)} the dynamical exponent for MN1 coincides with that of KPZ,
{\bf (ii)} the correlation length exponent, $\nu$
(that can be defined for MN but does not exist for KPZ, lacking of a critical point) 
can be related to the $z$ exponent in KPZ by $\nu = 1/(2 z -2 )$, and 
{\bf (iii)} the order-parameter critical exponent $\beta$ has to obey $\beta>1$.
The reader interested in the proof of these properties is deferred to \cite{MN2}
and  to \cite{Granada} for a more detailed explanation. 
   
\subsubsection{Large N limit}
  Exact RG calculations can be performed in the large N limit for the O(N)-symmetric
version of the MN equation defined as
\begin{equation}
   {\partial}_{t} \phi_\alpha(x,t) = - a \phi_\alpha -
   b |\phi|^2 \phi_\alpha + D {\nabla}^{2} \phi_\alpha(x,t) + 
 \sigma |\phi(x,t)| \eta_\alpha(x,t),
\label{MNN}
\end{equation}
where ${\bf \phi}(x,t) = \{ \phi_1, \phi_2, ..., \phi_N \}$ is an N-component
real vector field, and ${\bf \eta}$ is a noise vector field. 
The cubic term is necessary in order to preserve the $O(N)$ symmetry. The continuous 
symmetry of this model allows for two distinct types of active phases: one that
preserves the symmetry and one which breaks it. The first one has 
$\langle {\bf \phi } \rangle =0 $ and $Q = \langle  |\phi|^2/N \rangle \neq 0$, while 
for the second both order parameters are non-zero.
   The  main result of the exact solution presented in \cite{MN1} is that
for $d \le 2$, only the symmetric active phase exists, while for $d > 2 $
both types of active phase can exist: for large values of $\sigma^2$ the active phase
is symmetric, while a broken-symmetry state emerges below a certain critical
value of $\sigma^2$. There is also a multicritical point at  which the two active and the
absorbing  phase join (see \cite{MN1} for more details). At $d=2$ a {\it Kosterlitz-Thouless}
type of transition could appear \cite{KT}.

  \subsection{Microscopic models and Numerics} 

 Given the lack of analytical tools allowing for a determination of critical 
exponents in the strong coupling regime, one has to resort to numerical analysis.
The first estimation of critical properties was obtained in \cite{MN2} by discretizing 
Eq.(\ref{MN}) and performing a numerical integration in $d=1$. 
A more complete study was presented afterwards in \cite{MN4} 
in which the two- and three-dimensional cases were also analyzed.

In what respect to discrete, microscopic realizations of this universality class, amenable 
to be simulated, let us remark that any of the many discrete models in the KPZ class 
reported in the literature \cite{HZ,Krug,Laszlo} 
is, in principle, susceptible to be modified by introducing a limiting wall, 
which induces a phase transition. 
If the non-linearity sign (that can be numerically determined by performing 
velocity measurements of tilted interfaces \cite{Laszlo}) 
is positive (negative) an upper (lower) wall is required to construct a
model in the MN1 class.  
To the best  of our knowledge, at least three different discrete models with these
requirements have been studied in the literature \cite{MN3,Haye1,Ahlers}.


 In finite size analysis of Eq.(\ref{MN}) \footnote{
See \cite{Maxi} for a review of numerical 
methods designed to deal with stochastic equations.} one starts with 
{\it homogeneous initial conditions}, and
first determines $\langle \phi \rangle$ as a function of the system size, $L$,
for different control parameter values. In a double-logarithmic plot the critical
point corresponds to the straight line separating curves bending upward (pinned phase) 
from downward-bending ones (depinned phase). In order to have a well defined pinned
phase for small systems one has to restrict the statistics to pinned runs (otherwise, the 
only true stationary state for finite systems is the depinned or absorbing one 
\cite{Haye,MD,Granada}). 
In systems with MN one has to define a criterion to decide when a configuration is depinned.
Numerical results should not (and do not) depend on such a criterion.
From the slope of $\log(\phi)$ versus $\log(t)$, one extracts $\theta=\beta/\nu_\parallel$
(see Fig.(\ref{DP}) to have a glance at how finite size scaling graphs look like).
From the scaling of saturation values at criticality for different $L$ one
obtains $\beta/\nu_\perp$.
Considering a large system size it is possible to obtain $\beta$ from the scaling of 
saturation values at various distances from the critical point. In what respect 
the dynamical exponent, $z$, it can be determined either from studies of the 
two-point correlation function at criticality \cite{MN2} or from {\it spreading 
experiments} \cite{MN3,MN4}
\footnote{In spreading experiments $\phi(x)=0$ at
time $t=0$ at every single sites, except for a seed-site. See \cite{Haye,MD,Granada}.}.
Using spreading experiments at criticality the exponents $\delta$ and $\eta$,
associated with the surviving probability and evolution of the total number
of pinned sites respectively, can also be measured \cite{Haye,MD,Granada,Odor}.

In Table 1 we list the values reported so far in the literature for $d=1$. 
Even though there is some dispersion in the data, and a high precision
determination of critical exponents is still lacking, the existence of universal values
is firmly established.
\begin{table}[t]
\begin{tabular}{lcccccc}
  &  \multicolumn{6}{c}{} \\
  \cline{2-7}
  &                   $\beta$ & $\nu_\perp$ & $\beta/\nu_\perp$ & $z$ & $\theta$ & $\eta$ \\
  \hline
  MN  \cite{MN2}    &$1.70(05)$ & $1.03(05)$  & $1.65(10)$ & $1.53(07)$ & $1.10(05)$ & \\
  Hwa \cite{MN3}    &$1.50(15)$ & $\approx 1$ & $1.55(15)$ & $1.50(05)$ & $1.10(12)$ &\\
  MN  \cite{MN4}    &$1.50(05)$ & $1.0(1)$  &   $1.50(10)$ & $1.52(03)$ & $1.10(10)$ & -0.4(1)\\
  RSOS\cite{Ahlers} &$1.69$ & $$  &   $$ & $$ & $1.19$ &  \\
  Exact             &$ > 1$        & $ 1 $  &   $ $ & $1.5$ & & $$ \\
\hline
\end{tabular} 
\caption{Critical exponents measured for the MN1 class in $d=1$. 
RSOS stands for ``Restricted solid on solid'' model \cite{Ahlers}. 
MN stands for simulations performed directly on the Langevin equation}
\end{table}
\begin{table}[t]
\begin{tabular}{lcccccc}
  &  \multicolumn{6}{c}{} \\
  \cline{2-7}
  &              $\beta$ & $\nu_\perp$ & $\beta/\nu_\perp$ & $z$ & $\theta$ & $\eta$ \\
  \hline
Weak c.  \cite{MN4}&$0.97(05)$ & $0.50(5)$ &$1.94(2)$ & $2.00(0.5)$ & $1.0(1)$ & $-0.10(05)$ \\
Strong c.\cite{MN4}&$2.5(1)  $ &   $0.75(3)$ &$3.3(3)$ & $1.67(03)$ & $2.0(1)$ & $-0.5(1)$ \\
\hline 
Mean Field &$\beta_{mf}$& $ 1/2 $ &  $2  \beta_{mf} $ & $2$ &$\beta_{mf}$ & $0$\\  
\hline
  \end{tabular}
\caption{Critical exponents measured for the MN1 class in $d=3$, for both
the weak  and the strong coupling  regime  \cite{MN4}. Simulations were
 performed directly in the Langevin equation with $p=2$. 
The two different regimes are obtained by fixing the
noise amplitude to a small and a large value respectively; see \cite{MN4} for further details.
$\beta_{mf}$ is given by Eq.(\ref{mf}).}
 \end{table}
In $d=3$ the two different universality classes predicted by the RG calculation 
(see Fig.(\ref{flow})) can be easily recognized:
for small $\sigma$ values very close to mean field expectations have been reported 
\cite{MN4} while, for a larger $\sigma$ 
non trivial exponents have been found \cite{MN4} (see Table 2).

\subsection{Realizations}

   Among the physical realizations of this class, probably the most relevant 
one is a rather generic class of synchronization transitions in extended systems 
(see section \ref{synch}) \cite{PK,Lucia,Ahlers,Peter,death}.

 Another important application is in growing interfaces with both adsorption 
and evaporation of particles on a substrate. This may lead to pinning-depinning
transitions or to  non-equilibrium wetting transitions 
\cite{Haye1,Lisboa,Lisboa2,Lisboa3} (see \ref{NW}).
It might also be related to the {\it delocalization transition} in an 
adsorption-reaction model reported in \cite{Ayse}.

In reaction-diffusion systems, many times it is possible to derive a
Langevin equation in an exact way. In many cases, a multiplicative noise
appears, but many times it is purely imaginary! (see the last section
of this review ref{mix}, and the references cited there), leading  to
a completely different phenomenology.

\section{The MN2 class}
 
 As already anticipated keeping the KPZ non-linearity
to be positive, the presence of a lower wall leads to a MN equation including  
an extra term with respect to the previously  studied case: Eq.(\ref{MN2}).
It has a phenomenology substantially different from that of MN1,
Eq.(\ref{MN}).
In this section we underline the analogies and differences between these two cases 
by going through the different levels of approximation 
discussed in the previous section. 

\subsection{Zero dimensions}
The zero-dimensional case is obviously identical to the one for MN1.

\subsection{Mean Field approach}
 Concerning the mean field approximation, the average value of the
lower-wall extra term, $ -2 D{(\nabla \phi(x,t))^2 /  \phi(x,t)} $  generates
an exponential non-integrable singularity, $\propto \exp( + \langle \phi 
\rangle^2 / \phi^2)$,
at the origin of the quasi-stationary 
potential, solution of the associated Fokker-Planck equation, 
rendering it non-integrable. 
Only the depinned (absorbing) phase exists within this approximation.
A discussion of different types of mean-field approaches, all of them leading to 
this same conclusion can be found in \cite{Tafa}.

\subsection{Beyond Mean Field - Perturbative expansion}
 The additional term  $-2 D (\nabla \phi)^2  / \phi$ 
is non-analytical at $\phi=0$, rendering un-feasible any type of perturbative
expansion.
Renormalization group results analogous to those presented for the MN1 class are
therefore not available.
The arguments leading to  $\nu = 1/(2 z -2 )$ obtained for MN1 can be  extended to this
case, but the one leading to  $\beta > 1$ cannot. Indeed, as shown below, in this
case $\nu \approx 1$ as expected, and $\beta$ takes a value not far from $1/3$.

\subsection{Microscopic models and Numerics}
In what respects numerical results, direct simulations of Eq.(\ref{MN2}) 
are hopeless owing to the presence of the singular term. 
Contrarily, simulations performed in the KPZ representation 
are feasible, although they are also hard-to-deal-with as is generically the case
when discretizing KPZ-type of equations \cite{KPZproblem}.

 At the light of this barren perspective, one has to resort to simulations
of microscopic models. The first study was that of \cite{MN3}, where
critical exponent close to those of the zero-dimensional case were reported,
using  a well known KPZ-like model \cite{MN3,Krug}.
However, it has been recently shown that the results reported in \cite{MN3} 
correspond to effective exponent values computed in a transient regime
\footnote{Apart from the numerical estimates, the rest of the discussion presented 
in \cite{MN3} about the differences between the upper and the lower wall 
cases remains valid.}, 
and new accurate values have been measured (results are listed in Table 3 \cite{Tafa}).
These studies are performed: ({\bf{i}}) using a restricted solid-on-solid (RSOS) model
\cite{HZ,Laszlo,Krug,Ahlers} known to  belong, in the absence of a wall,
to the KPZ class, and reaching system sizes as large as $2^{20}$,
and ({\bf{ii}}) the model considered in \cite{MN3} but  with much larger statistics 
and performing finite size scaling analyses
\footnote{The discrete model considered in \cite{Haye1} is 
extremely  slow to simulate, as exhibits manifestly severe crossover effects.}.
Results for both models coincide within errorbars as shown in Table 3.
\begin{table}[t]
\begin{tabular}{lcccccc}
  &  \multicolumn{6}{c}{} \\
  \cline{2-7}
                & $\beta$ & $\nu_\perp$ & $\beta/\nu_\perp$ & $z$ &
$\theta$ & $\eta$ \\
  \hline
 Hwa \cite{Tafa}  &$0.32(3)$& $0.97(5)$  & $0.33(2)$ & $1.55(5)$ &
$0.215(15)$ & $$\\
 RSOS \cite{Tafa} &$0.325(5)$& $\approx 1$  & $0.33(2)$ & $\approx 1.51$ &
$0.21(1)$ & $0.80(2)$ \\
 \hline
   \end{tabular}
\caption{Critical exponents for the  MN2 class in $d=1$, as measured in two different models:
model 1 stands for the one described in \cite{MN3}, and Model 2 is a RSOS one. See
\cite{Tafa} for more details.}
\end{table}

These values are clearly different from those reported for the upper wall, confirming
once again the dramatic (ever-surprising) difference between the two classes, as well
as the existence of the robust MN2 class itself. 

The situation is much less satisfactory  here than it is for the MN1 class.
We have found problems both in mean field and in field-theoretical approaches, and
even the numerical integration of the Langevin equation, in either the density  of
the interface representation is problematic \cite{Tafa}
This calls for the development of new analytical tools aimed at tackling 
this problem and clarifying the physics of this (well identified and
robust) universality class.

\subsection{Realizations}
 
 The problem of probing the alignment of different DNA sequences is of outmost 
importance in molecular biology. One of the most widely used algorithms for such
a problem is due to Waterman \cite{Hwa}. 
Hwa and collaborators, showed that such an algorithm can 
be mapped onto a KPZ problem with a lower wall (in particular, to its
DPRM counterpart) \cite{Hwa}. Locating the critical point of such a transition is
what is effectively done by molecular biologists 
when tuning parameters (\ie  choosing a {\it scoring scheme})
to enhance or maximize fidelity in local alignment studies  \cite{Hwa}.

 The MN2 class appear also in some pinning-depinning or non-equilibrium 
wetting transitions in which a KPZ interface (with positive non-linearity) 
grows from a substrate \cite{Haye1,MN3}.

 It is known that one-variable MN processes under various 
circumstances (as, for example, the presence of {\it reset} events) lead to generic
scale invariance \cite{Sornette,Solomon,Manrubia}. The MN2 equation is the natural extension
of this phenomenon to extended systems, and has been proposed, for instance, to describe
the properties of {\it city growth} \cite{Zhang,Manrubia}, where fluctuations in 
population number are multiplicative and, obviously, bounded from below.

\section{Changing the potential: MN1 with an attractive wall
\label{wet}}

\begin{figure}
\begin{center}
\includegraphics[width=0.3\textwidth,height=0.4\textheight,angle=-90]{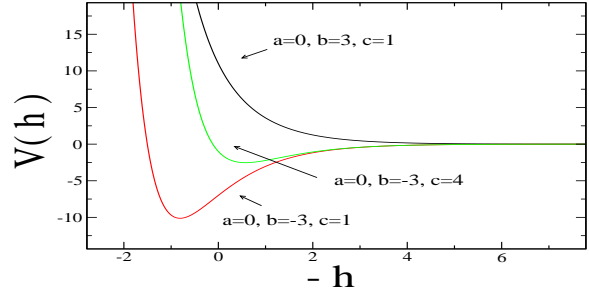}
\caption{Effective potential in terms of $h$ for different parameter values. 
For $b<0$ it has a minimum nearby the wall.}
\label{potential}
\end{center}
\end{figure}

Up to now we have dealt with Langevin equations describing continuous
transitions in systems with MN. 
In the same way as the Model A for Ising like transitions Eq.(\ref{ModelA}) has to  be
modified to the following Langevin equation
\begin{equation}
   {\partial}_{t} \phi(x,t) = - a \phi -
   b \phi^3 - c \phi^5 + D {\nabla}^{2} \phi(x,t) +  \sigma \eta(x,t).
\label{ModelA2}
\end{equation}
with $b<0$ and $c>0$, in order to describe discontinuous, 
first-order, transitions (in the absence of external magnetic fields)
one might wonder what is the MN version of Eq.(\ref{ModelA2}) and what 
type of physical problems could it represent.
To explore this possibility one just needs to change the potential in Eq.(\ref{MN})
to allow for discontinuous jumps:
\begin{eqnarray}
{\partial}_{t} \phi(x,t) & = & - {\delta V(\phi)  \over \delta \phi} 
+ D {\nabla}^{2} \phi(x,t) + \sigma  \phi(x,t) \eta(x,t)  \nonumber \\
V(\phi) &= & {a \over 2} \phi^2 + {b \over 3} \phi^3 +  { c \over 4} \phi^4
\label{Wet}
\end{eqnarray}
with $b<0$ and $c >0$. 
If $\phi \leftrightarrow -\phi$ symmetry is to be respected, the simplest potential
becomes $ a \phi^2 + b \phi^4 + c \phi^6$ (with no essential difference in the 
underlying physics). 
Transforming Eq.(\ref{Wet}) to the interface language:
\begin{eqnarray}
\partial_t h(x,t) & = & -{\partial V(h) \over \partial h} +
 D \nabla^2 h + D (\nabla h)^2 + \sigma \eta(x,t) \nonumber \\
V(h) &= & a~ h + b ~e^h + c/2 ~e^{2 h}.
\label{KPZS}
\end{eqnarray}
For $b<0$ this is a KPZ equation in the presence of an {\it attractive upper wall}, \ie
in order to depin a site close to the wall, a certain potential barrier
has to be overcome. Indeed, the potential $V(h)$ has a local minimum nearby the wall
(see Fig.(\ref{potential}).
In analogy with the previous sections, this KPZ equation with a {\it positive non-linearity} 
and an upper wall is fully equivalent to a negative non-linearity equation with an
attracting lower wall.  
Eq.(\ref{Wet}) is therefore the $b<0$ counterpart of the MN1 class.
 This equation has been studied in different contexts in the literature.
It was first proposed by  M\"uller \ea \cite{Muller} to study first order
non-equilibrium transitions in extended systems. Afterwards, a very similar equation
was proposed by Zimmerman \ea \cite{Mallorca} as a simplified stochastic model 
of {\it spatio-temporal intermittency}, and by Giada and Marsili \cite{Marsili} to study 
the effect of {\it Morse potentials} in KPZ dynamics 
\footnote{Observe that $V(h)$ has the form of a Morse potential.}.
Also, in an interesting paper, Hinrichsen \ea \cite{Haye2} studied what they called 
{\it first order non-equilibrium} wetting: a discrete random 
deposition model, known to be described by KPZ, 
in the presence of both evaporation ({\ie a drift) and a limiting 
attractive substrate (\ie the motion of the KPZ interface
is not just limited by the wall, but the wall is ``sticky'' in the sense that sites 
attached to the wall need to overcome an extra potential-barrier in order to become 
depinned) was analyzed. 
It can be easily argued \cite{Haye2} that this discrete model
is represented at a continuous level by Eq.(\ref{Wet}).

Following the same steps as in preceeding systems we now analyze Eq.(\ref{Wet})
employing different methods.

\subsection{Zero dimensions}
In zero dimensions, the problem is exactly solvable. The stationary 
probability distribution being (see appendix B)
$P_{st}(\phi)= {1 \over \phi^{1+2a/\sigma^2}} 
e^{{-2 \over \sigma^2} \left( b \phi +{c \phi^2 \over 2} \right)}.$
At the transition the order parameter jumps discontinuously from 
$0$ to a non-vanishing value.

\subsection{Mean Field approach}
 At a mean field level, we can perform a calculation analogous to that
presented for Eq.(\ref{MN}). 
The numerical solution of the corresponding self-consistency  equation generates 
the phase diagram shown in Figure (\ref{mf-fig}). We describe now the main
features 
(for a more detailed explanation see \cite{Muller,Marsili,Lisboa,Lisboa2}).
\begin{figure}
\begin{center}
\includegraphics[width=0.5\textwidth,height=0.2\textheight]{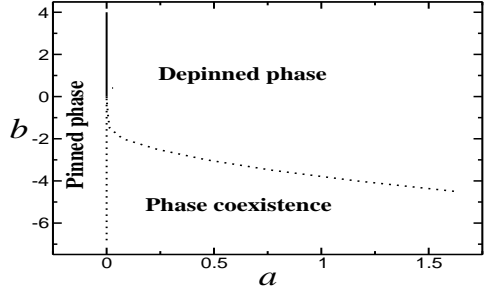}
\caption{Phase diagram derived from the numerical integration of the 
mean field approximation of Eq.(\ref{Wet}). In the coexistence regions both, the
pinned and the depinned solutions are stable, and their edges (marked by dashed lines)
correspond to discontinuous transitions, \ie one of the phases loses its stability  abruptly.
The transition corresponding to the upper one,  $a_c(b)$ (signaling the limit of
stability of the pinned phase), may become continuous once fluctuations
are considered, as discussed in the text.}
\label{mf-fig}
\end{center}
\end{figure}
For positive values of $b$ the transition occurs at $a=0$, it is continuous and
very similar to the one in Eq.(\ref{MN}): it is in the MN1 class 
\footnote{As in MN1, in this case,
for $a>0$ ($a<0$) the tail of the potential would attract (repel) distant
interfaces to (away from) the wall; the transition from one case to the other
occurs at $a=0$.}.
Instead, for $b<0$ the transition bifurcates: there is a discontinuous transition at
$a=0$, and a second transition (also discontinuous at mean field level)
appears at a positive value of $a$ (see Fig.(\ref{mf-fig})),  $a_c(b)$.
At this value, 
the pinned phase, $\langle \phi \rangle  \neq 0$, loses its stability
and falls in an abrupt way to the depinned phase $\phi=0$.
The depinned phase loses its stability at a smaller value $a^*=0$.
Therefore, in the interval $[0,a_c(b)]$ both solutions are stable:
we are in the presence of {\it phase coexistence}.  The existence of broad coexistence
regions is a trait specific of non-equilibrium phase order transitions, and have
been reported to appear, among others, in the Toom model \cite{Toom}. 
Observe that both of the limiting lines of the coexistence region are
discontinuous transitions at this level.

 \subsection{Beyond Mean Field - Perturbative expansion}
 The theory can be treated perturbatively in a very similar way to Eq.(\ref{MN});
indeed the Feynman diagrams have the same topology in both cases.
The new constant $c$ renormalizes as 
 $ c_R  =  c  (1 +   \sigma^2 I + \sigma^3 I^ 2  + ...)=   {c  \over 1 - \sigma^2 I}$
and a RG flow analogous to that for Eq.(\ref{MN}) is obtained. 
The only  difference is that now as long as the transition is discontinuous (not
scale invariant) the trajectories for large enough noise amplitudes 
(above the separatrix in Fig.(\ref{flow}))
are not expected to flow to a strong coupling fixed point, but to be runaway 
trajectories. 
In any case, there is no non-trivial stable fixed-point and, consequently, standard 
perturbative expansions cannot say anything about the strong noise regime.   
\subsection{Microscopic models and Numerics}
 Eq.(\ref{Wet}) has been studied numerically in different contexts \cite{Mallorca,Marsili}
and recently in \cite{Lisboa,Lisboa2}. Its main properties (described mainly in the interface
language) are as follows. 
The 
mean-field phase-diagram qualitative-structure survives the inclusion of fluctuations,
 namely:
\subsubsection{$b>0$}
 For $b>0$ a continuous phase transition is observed. 
This transition has been extensively verified to belong to the MN1 class \cite{Lisboa,Lisboa2}.
\subsubsection{$b<0$}
 For $b<0$ simulations performed taking $D=1$ confirmed the existence of a broad coexistence 
region. Within this region an initially  detached interface falls
towards minus infinity with a finite average velocity (\ie absorbing  remains absorbing)
while an initially pinned region remains pinned. This is strictly true only in the 
thermodynamic limit.
For finite sizes, the only stable state is 
the depinned (absorbing) one \cite{Haye,Granada}.
 The characteristic time required for an initially pinned interface to detach 
goes to infinity for large system 
sizes within the coexistence region, confirming the thermodynamic stability of the 
pinned phase.
Deep into the pinned (active) phase this characteristic time grows exponentially,
while close to $a_c(b)$ there is a tiny region where such a time seems 
to grow as a power law 
\footnote{In the absence of a theoretical understanding, it could
well be the case, that this power law is just a transient, becoming asymptotically 
an exponential. Indeed this seems to be the most likely scenario.}.   
Within this regime, {\it spatio-temporal coexistence} of the two phases is actually 
observed in computer simulations 
\footnote{Noteworthy that phase coexistence (between a pinned and a depinned phase)
is one thing, and spatio-temporal coexistence of detached and non-detached patches
is a different one. This last is usually identified with spatio-temporal intermittency.}.
 In Fig. (\ref{STI}) we see a space-time plot of the activity in this regime. 
Pinned and depinned regions coexist implying that there are stable mechanisms 
for both: generating fluctuations and eliminating islands 
of the minority phase. In this way a spatio-temporal intermittency (STI) regime 
\cite{Chate} is shown to exists \cite{Mallorca,Lisboa}. 
Analogous mechanisms, required to sustain two-phase
spatio-temporal structures, have been described in detail in a microscopic 
realization of this class \cite{Haye2,HayePCPD}.
\begin{figure}
\begin{center}
\includegraphics[width=0.5\textwidth,height=0.2\textheight]{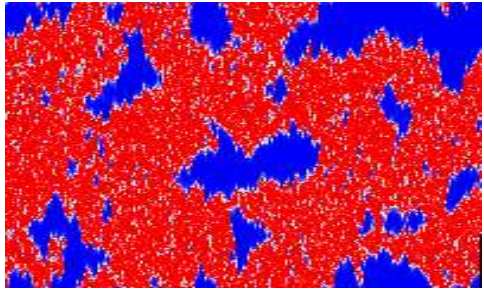}
\caption{Spatio-temporal evolution of Eq.(\ref{Wet}) in the spatio-temporal
intermittency region, close to $a_c(b)$. 
In the horizontal axis we plot the time, while the space-position is plot 
in the vertical one. Red (light) for pinned
sites, \ie large (small) $\phi$, and blue (dark) for depinned ones.}
\label{STI}
\end{center}
\end{figure}
Performing a cut for a given time, and plotting $h=-\log(\phi)$ as a function of the 
position, we obtain a profile as the one shown in Fig. (\ref{triangles}).
The depinned regions tend to escape from the wall, but being pinned at their
edges, they are constrained to form very characteristic {\it triangular patterns.}
\begin{figure}
\begin{center}
\includegraphics[width=0.35\textwidth,height=0.40\textheight,angle=-90]{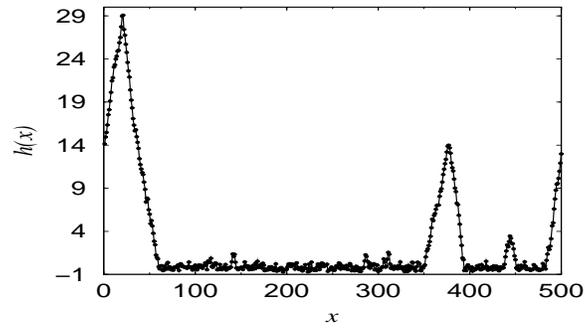}
\caption{Values of $h=-\ln(\phi)$ in the previous figure, at a given fixed time-slice
(roughly speaking at $3/4$ of the maximum time; the large triangle, corresponds to the
up-right absorbing region in the previous picture).}
\label{triangles}
\end{center}
\end{figure}
The characteristic slope, $s$ of the triangles can be determined by observing that
$\lambda  \langle (\nabla h)^2 \rangle$ has to compensate the effect of the 
constant drift $a$, and therefore $\lambda s^2=a$. 
As $\lambda$ can be measured from
the average speed of tilted interfaces \cite{Laszlo,HZ}, $s$ can be inferred.
Indeed, the  value obtained in this way is in very good agreement with the measured 
average slope as reported in \cite{Lisboa2,Korea}.
Summing up, inside the coexistence region we have two different sub-phases.
In both of them the depinned phase is stable, 
but:
\begin{enumerate}
\item  One has small fluctuations of the pinned phase, and exponential increase
of detaching times
Two homogeneous phases, pinned and depinned, coexist. 
\item The other has larger fluctuations in the pinned phase,
allowing for finite patches to depin and form typical triangular structures. 
In this regime {\it spatio-temporal intermittency} is observed.
The detaching times in this regime seem to grow algebraically  with $L$
although this could be a transient effect.
In this regime a STI-pinned phase coexists with the homogeneous depinned one. 
\end{enumerate}
\begin{figure}
\begin{center}
\includegraphics[width=0.5\textwidth,height=0.38\textheight]{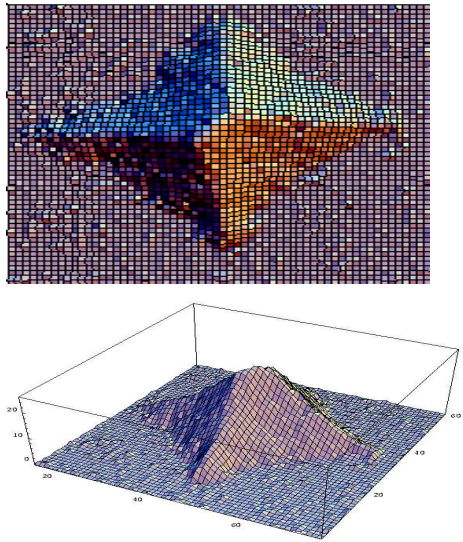}
\caption{Two views of a pyramidal structure appearing in two-dimensional 
simulations, in the phase coexistence region, of the MN noise equation in the
presence of an attractive wall.}
\label{pyr}
\end{center}
\end{figure}
  In higher dimensional systems the role of the triangles is played by {\it pyramidal structures}
(see Fig. \ref{pyr} and \cite{Lisboa2}.)

\subsubsection{A DP transition}
 Surprisingly enough, numerical studies have recently shown that there is a region
in parameter space for $b<0$, not predicted by the mean field approach, in which
the transition at $a_c(b)$ becomes a continuous one.
We will illustrate in what follows that the presence of fluctuations 
may induce the mean-field first-order transition at which the pinned phase looses 
its stability to become second-order, governed by DP exponents.
This is somehow surprising, as the broad coexistence region survives, but its upper
border edge corresponds to a continuous transition. \footnote{A similar situation (second
order transitions with first-order traits)
named ``{\bf one-way hysteresis}'', has been recently reported to occur in some 
different contexts. 
See R. Maimon and J. M. Schwarz, condmat/0301495 and references therein.}

In \cite{synchro} it has been shown that, at least, 
for certain parameter values,
satisfying  the following two requirements:
i) to generate at mean field level a
{\it weak } (small jump) discontinuous transition (in such a way that 
fluctuations can alter 
more easily its degree), and
ii) to have a relatively small coupling constant $D$, so the mechanism at the
basis of the instability of the pinned phase is weakened,
 \footnote{Detached triangles pulling
with a strength which depends on $D$ are responsible for the destabilization
of the pinned phase \cite{synchro,Haye2}
(See \cite{HayePCPD} for a nice explanation of the triangles dynamics 
and stability.} a continuous DP transition is
generated. The detailed phase diagram has still to be worked out and the conditions
under which a continuous transition is generated need to be further clarified. 

What is already firmly established from numerical simulations in \cite{synchro}
is that such a continuous-transition regime exists and that it is controlled 
by DP exponents.
In particular, in Figure \ref{DP} we show the order parameter $\langle \phi \rangle$
time  evolution for different system sizes. Both the slope at criticality  and the 
scaling of saturation values for finite sizes, show unambiguously DP values 
(see figure caption).
\begin{figure}
\begin{center}
\includegraphics[width=0.45\textwidth,height=0.25\textheight,angle=0]{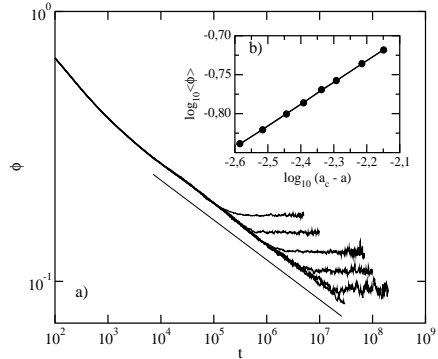}
\caption{Log-log plot of the order parameter time evolution (averaged over
surviving, pinned, runs) at the critical point, for system sizes
(top to bottom) $L=25, 50, 100, 200$, $400$, and $1000$.  Both the
main slope, $\theta=0.159(5)$, and the scaling exponent of saturation values,
$\beta/\nu_\perp = 0.245(15)$ reveal DP-scaling behavior ($\theta=0.15947...$ and
$\beta/\nu_\perp =0.25208(5)$ \cite{DPvalues}).
Inset: Log-log plot of
the average order parameter as a function of $|a_c -a|$. The full line
is a fit $\langle \phi \rangle \sim |a_c - a|^\beta$, with an exponent
$\beta=0.28(1)$, also compatible with the DP value $\beta=0.27649(4)$.}
\label{DP}
 \end{center}
\end{figure}
 In order to fully verify the presence of DP scaling, different tests have been
performed in  \cite{synchro}. 
In particular, the moment ratios have the same (universal) values as those measured
for the DP-class \cite{moments}. It was also verified that the number of sites
pinned by the potential scale with the same (DP) exponents. Finally, by computing
numerically the fluctuations from different initial states, it was checked that
they are proportional to the square-root of pinned (active) sites, as occurs in
the Reggeon-field theory describing  DP \cite{Haye,Granada}. 

 To further confirm the emergence of DP scaling with $b<0$, 
and in order to confirm the generality of the described
mechanisms, a well-known discrete interface model known
to be in the KPZ class (\ie the one described and studied in  \cite{MN3})
 has been studied by including an attractive wall.
The introduction of attractiveness changes in a very abrupt way the critical
behavior: while in the absence of attraction (purely bounding  wall) 
the transition is described by the MN1 class, as soon as attractiveness 
is switched on (and therefore $b$ changed from positive to negative)
the transition crosses over to DP. Combining Monte Carlo simulations
with finite size scaling, many decades of clean DP scaling can
be observed.

According to a recent preprint by Droz and Lipowski \cite{DL}, given that:
i) the absorbing (depinned) state is compatible with multiple possible
realizations, and ii) at the light of numerical simulations,
this transition is in the {\it DP class with infinitely many absorbing states}
\cite{IAS}. This means in particular that the spreading  exponents may be not
universal \cite{IAS}.

 Before ending this section, let us remark that it could well be the case, 
that this transition is controlled asymptotically, in $d=1$, by a DP fixed 
point for any parameter values.  
Indeed, some heuristic arguments have been proposed by Hinrichsen pointing out
to the impossibility to have true first-order transitions between an absorbing  
and a fluctuating phase in one-dimensional systems \cite{first?}. 
If this scenario is true, then {\it the apparent 
discontinuous transition at the upper border line of the coexistence region would be 
just a transient, crossing over to DP}. 
In this scenario, the described STI phenomenology would corresponds to DP-dynamics.
In higher dimensions, first order
transitions can appear without troubles. 
 
\subsection{Open problems}
An open problem is how to deduce analytically (\ie from a RG perspective) 
that Eq.(\ref{Wet}) can generate DP scaling for some parameter values.

An intuitive, heuristic justification of this fact is as follows: 
the scaling, dominated in general by interface fluctuations 
is controlled here (in the DP regime) by fluctuations of the number of pinned sites. 
The attractive wall, by generating a minimum in the potential, provides a physical 
ground to distinguish between pinned and not-pinned sites. We can assign them,
$1's$ or $0's$ respectively, and define in such a way an effective dynamics for
these booleans variables. This 
has to be very similar to DP. For instance, depinned (absorbing) regions will not become 
pinned (active) as, once away from the potential well, the interface 
is pushed exponentially towards infinity, so $1$ can go to $0$, but the reverse is not 
true.
However, from a more rigorous point of view, it has not been possible so far
to derive DP-scaling from Eq.(\ref{Wet}) using RG techniques. Indeed, as anticipated
previously perturbative studies can say nothing about the strong-noise
regime allegedly controlled by DP-scaling \footnote{
A very similar problem emerges in the context of fluctuating (KPZ) interfaces 
in the presence of quenched disorder. Also in that case, DP-scaling emerges
in a way unaccessible to standard perturbative methods (see Appendix C).}.

 Another problem is related to the possibility raised by Hinrichsen \cite{first?}
that the first-order transition at which the pinned phase becomes unstable, 
is only apparently discontinuous. It might well be the 
case that  the line of transition $a_c(b)$ is continuous all the way, 
with asymptotic DP-scaling. This needs to be further clarified in more
extensive computer simulations.

\subsection{Realizations}
 
  The main two realizations known so far of this type of MN equation in the 
presence of an attractive wall are
non-equilibrium wetting and synchronization in extended systems.
These two realizations are discussed in detail in  forthcoming sections
  
   It could also be related to the Pair Contact process with
diffusion (one of the most intensely studied and debated model with 
absorbing  states \cite{PCPD}) for which  a similar
equation with MN has been derived \cite{Tauber,nplet}, as pointed out very recently
by Hinrichsen \cite{HayePCPD}. This could justify the conclusion that
this controversial model exhibits asymptotically DP scaling.

\begin{figure}
\begin{center}
\includegraphics[width=1.0\textwidth,height=0.25\textheight,angle=0]{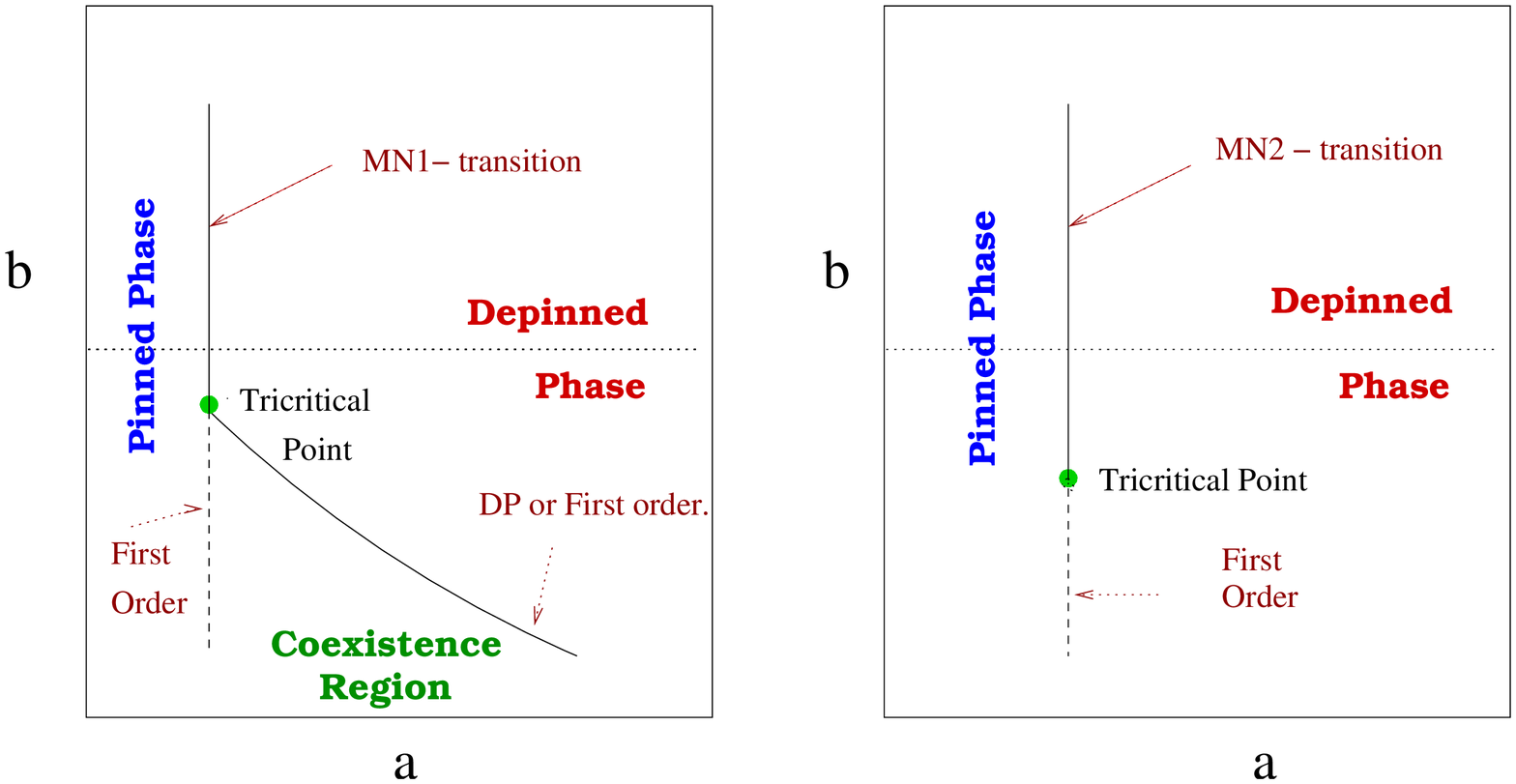}
\caption{Schematic representation of the phase diagram for upper (lower) and lower 
(upper)
walls with a positive (negative) non-linearity. See the caption of Table 4}
\label{diag}
\end{center}
\end{figure}

\begin{table}
\begin{center}
\begin{Large}
\begin{tabular}{|c|c||c|c|}
  \multicolumn{4}{c}{} \\
  \hline
           &  & Upper Wall & Lower Wall   \\
  \hline
   \hline
   NO & $\lambda > 0$  & {\it MN1} & {\it MN2} \\
\hline 
 NO   & $\lambda < 0$  & {\it MN2} & {\it MN1} \\
\hline 
 YES & $\lambda > 0$  & BPC: $1^{st}$ or DP &  {\it MN2~ or $1^{st}$  } \\
\hline
 YES & $\lambda < 0$  & {\it MN2~ or $1^{st}$}   & BPC: $1^{st}$  ~ or ~ DP \\
\hline
  \end{tabular}
\end{Large}
\end{center}
\caption{Universality  classes for the KPZ equation
with different signs of $\lambda$, in the presence of
Upper or Lower walls, with ({\bf{YES}}) or  without ({\bf{NO}}) attractiveness.
 {\bf MN1} stands for the universality class described by Eq.(\ref{MN}); 
 {\bf MN2}  stands for Eq.(\ref{MN2}); 
 {\bf  $1^{st}$}  represents discontinuous transitions,
 {\bf DP} transitions in the directed percolation class, and
 {\bf BPC} for a broad region of phase coexistence. 
In the case of a BPC region there are two phase transitions, one 
(where the depinned phase looses its stability) is always discontinuous
(not reported in the table)
and another (at which the pinned phase becomes unstable) which may be either
first order (in dimensions larger than two, and probably in an apparent, 
transient way in $d=1$) or DP-like, as shown in the table. Also, in this
case, if the value of $|b|$ is small enough, the renormalized value of $b$
might be positive leading to a standard continuous MN1 transition, without BPC.
In the case of MN2 with attractiveness no BPC appear; the transition is second
order, in the MN2 class, except for large negatives values of $b$ where 
it becomes discontinuous.} 
\end{table}

\section{The remaining possibility: MN2 with an attractive wall}

The equation described in the previous section is, as explained, the 
$b<0$ (attractive wall) counterpart of the MN1 class. To exhaust all
the possibilities, we just need to explore the phenomenology of
attractive walls ($b<0$) added to the MN2 class, \ie lower attractive
walls with positive non-linearities or equivalently upper attractive
walls with a negative KPZ non-linear coefficient 
\footnote{We have to admit that this looks like a tongue-twister and we humbly 
suggest the long suffering reader to draw a diagram to clarify the list of
possibilities. 
This diagram should include $2^3$ situations:
upper or lower wall?, 
positive or negative non-linearity?, attractive or non attractive wall?; and
only  4 different physical behaviors, described respectively by this and the 3 
preceeding sections. Table 4 and Fig. (\ref{diag}) should be of some help.}.

  In the interface language we have
\begin{eqnarray}
\partial_t h(x,t) & = & -{\partial V(h) \over \partial h} +
 D \nabla^2 h - D (\nabla h)^2 + \sigma \eta(x,t) \nonumber \\
V(h) &= & a~ h + b ~\exp{h} + c/2 ~\exp(2 h) 
\label{KK}
\end{eqnarray}
with $b<0$. This can be mapped into
\begin{equation}
{\partial}_{t} \phi(x,t) = - a \phi -
b \phi^2 - c \phi^3 + D {\nabla}^{2} \phi(x,t) 
- 2 D {(\nabla \phi(x,t))^2 \over  \phi(x,t)} + 
\sigma  \phi(x,t) \eta(x,t).
\label{MN2A}
\end{equation}
with a negative value of $b$.
Having arrived almost exhausted to this fourth possibility we
will not split this section into the usual subsections of the preceeding ones. 
We just stress that $b<0$ does not have effects as dramatic as it had in the
previous case. Here  the transition,
which at mean field level is discontinuous, becomes continuous for relatively small
values of $|b|$ 
(and belongs to the same universality class as the $b>0$ case \ie to the MN2 class.),
while for sufficiently large ones, it remains first order without a broad 
coexistence region 
\footnote{Observe that the first-order transition 
here does not require the presence of coexistence 
between active and absorbing  phases: it just  corresponds to the point where the
free interface has zero velocity, and therefore does nor contradict Hinrichsen's claim
\cite{first?}.}.
 In this way,
the presence of fluctuations may drive the transition from
discontinuous (at mean field level) to continuous in some parameter range.
In other words the tricritical point is
shifted from $b=0$ at mean field, to a negative value when fluctuations are taken
into  account (this effect is much stronger here than it  is in the previous case).
This result has been verified in numerical simulations both using the model in
\cite{MN3} and a RSOS model in the presence of an attractive wall \cite{Tafa}. 
Let us stress again that even if the transition is discontinuous, first-order, 
in the strong attraction regime, it does not involve a broad coexistence 
region nor a region of  STI as occurred
in the previously  discussed case. 
The transition is located for any value of $b$ at the
point where the depinned phase (free interface) looses its stability.        

 One way to rationalize this result (and its being at odd with the previous section
one) is as follows. In the MN1 class in which 
there is an effective repulsion from the wall as discussed at the beginning of the
paper, the introduction of attractiveness has a profound effect. In particular, 
two opposite physical mechanisms compete: 
one repelling sites from the wall and the other attracting  
them to it. This is at the origin of the existence of a broad coexistence region.
Contrarily, for the MN2 class such an effective repulsion does not exist and,
therefore, introducing  attractiveness is a much milder perturbation in this 
case. Indeed, it does not alter nor the order nor the exponents of the
transition that remains in the MN2 class for sufficiently small values of $|b|$, and
it induces a standard discontinuous transition (no broad coexistence region) 
for large enough (very strong) attractiveness. 

 A similar phase diagram, including first and second order transitions as well
as a tricritical point separating them, has been reported by Hinrichsen \ea in their
discrete non-equilibrium wetting model, for parameter values corresponding to the
case under consideration.  

\section{Applications: Non-equilibrium wetting and synchronization \label{App}}

  In this section we briefly review two of the main fields in which the
formalism and results of the previous sections can be straightforwardly 
applied: non-equilibrium wetting and synchronization in spatially extended systems.
We discuss how MN equations as those described in the  preceeding  sections
come out within these contexts.

\subsection{Non-equilibrium Wetting \label{NW}}

When a phase ($\alpha$) is in contact with a substrate,
{\it wetting} occurs if a macroscopic layer of a coexisting
phase ($\beta$) is adsorbed at the substrate.
The wetting transition is characterized by the divergence of the
$\beta$-layer thickness; the $\alpha \beta$ interface goes arbitrarily far from the 
substrate.
Equilibrium wetting is a problem of great experimental as well as theoretical
 importance. It
has been extensively investigated
using interface displacement models, in which $h(x,t)$ represents the 
interface distance from the substrate \cite{Rev-wet}.
The minimal dynamical (equilibrium) wetting model 
is given by the following Edwards-Wilkinson (EW) \cite{EW} Langevin equation
\cite{EW+wall}
\begin{equation}
\partial_t h({\bf x},t) =
D \nabla^2h -{\partial V \over \partial h} + \sigma \eta({\bf x},t).
\label{eweq}
\end{equation}
This was introduced by Lipowsky \cite{EW+wall}, and describes the
relaxation of $h$ towards its equilibrium distribution, specified by 
$V(h)$. A physically motivated choice for the interaction potential has
the following form (Morse potential) \cite{Rev-wet}
\begin{equation}
 V(h)=b(T)e^{-h}+ce^{-2h}
\end{equation}
where $b(T)$ vanishes at the wetting temperature and $c>0$.
At sufficiently low temperatures, $b<0$, the potential $V$ binds the
$\alpha \beta$ interface; the equilibrium thickness of the
wetting layer $\langle h \rangle$ is finite.
As the temperature is raised, the potential becomes less attractive and
at some point it no longer binds the interface, $\langle h \rangle$ diverges.

In order to study {\it complete wetting} (\ie wetting occurring in the limit 
in which the chemical potential difference between the two phases, $\mu$, 
goes to zero for $T$ larger than the wetting-temperature) 
a linear term $\mu h$ has to be added to $V(h)$ \cite{Rev-wet}.
 $\mu$ plays the role of an external force acting on the interface.

The dynamic model, Eq.(\ref{eweq}) is readily generalized to
non-equilibrium interfacial processes, {\em e.g.\/}, crystal growth,
atomic beam epitaxy, etc., where thermal equilibrium does not apply.
This has been done in detail in \cite{Lisboa,Lisboa2,Lisboa3}.
\footnote{See also \cite{Haye1,Haye2} where the same problem was first
studied for discrete models.}
 The conclusion is that the most basic 
effective non-equilibrium interfacial model consists of a
KPZ equation (instead of Eq.(\ref{eweq})) 
in the presence of a substrate, {\it i.e.:} 
the MN equation. The origin and nature of the
KPZ non-linearity within this context has been discussed in \cite{Lisboa,Lisboa2,Lisboa3}. 

A detailed discussion about non-equilibrium critical wetting,
complete wetting, the difference between pinning-depinning and wetting-dewetting
transitions, and other related issues can be found in \cite{Lisboa2,Lisboa3} (see also 
\cite{last}). 

\subsection{Synchronization \label{synch}}
 Mutual {\it synchronization} of oscillators is one of the most
intriguing and fascinating phenomena appearing in complex systems  \cite{Strogatz,Syn}.
Some examples appear in
chemical reactions \cite{chemical}, 
neuronal networks \cite{neurons},
flashing fireflies \cite{fireflies},
Josephson junctions \cite{Josephson},
and semiconductor lasers \cite{lasers} to mention but a few.
See \cite{Strogatz} for an elementary introduction
to this growing field and \cite{Syn} for an excellent review.

Coupled map lattices (CML) \cite{Kaneko,CM} have
attracted much attention as a paradigm
in the study of synchronization in spatially extended
systems under mathematical grounds.
When two replicas of a same CML are {\it locally} coupled
\cite{Peter,Zanette}, or when they are coupled to a 
sufficiently large common external random
noise even if they are not coupled to each other
\cite{firenze1,first,Lucia}, they can achieve mutual synchronization. 

 In all the aforementioned examples, there is a transition from a 
chaotic or unsynchronized phase, in which the differences between both
replicas persist, 
to a synchronized one in which memory of the initial differences is
erased and replicas synchronize with certainty.
The analysis of universal properties of 
synchronization transitions (ST) has been the subject of many recent studies
\cite{Peter,death,Lucia,PK,Ahlers}.
In particular, as mentioned previously, in a seminal paper, Pikovsky and Kurths 
devised a method to map the synchronization error field (\ie the difference between 
two replicas) into the MN equation \ref{MN}. On the other hand, 
it was proposed that for discrete systems the ST should be in the 
DP class \cite{Peter,death}. 

Based on recent observations \cite{Lucia,Ahlers,first} 
a global picture, that can be synthesized as follows, has emerged:
there are two types of ST depending on whether the largest (transverse) 
Lyapunov exponent, $\Lambda$, is zero or negative right at the 
transition \cite{Lucia,Ahlers}.
 In the first case, the transition is MN like, and it is controlled by linear 
stability effects ({\ie $\Lambda$ changes sign at the transition).
In the second group the transition is either DP or discontinuous depending 
on microscopic details (and the transition is located where the propagation 
velocity of non-linear perturbations changes sign, while $\Lambda$ is
negative). 

   This situation is in complete analogy with what reported for Eq.(\ref{Wet}).
Indeed, as shown in  \cite{synchro}, 
a straightforward generalization of the derivation by Pikovsky and 
Kurths leads to Eq.(\ref{Wet}) instead of Eq.(\ref{MN}) 
in the cases when there is some singularity or discontinuity in the local map.
\footnote{In a recent paper \cite{Ginelli} it has been shown how DP can 
emerge in synchronization transitions, even for some continuous maps.}
In particular, as discussed in \cite{synchro} if there is a typical re-injection 
rule or another mechanism leading to a typical value of the local 
synchronization error field, then the corresponding potential in the
Langevin equation needs to have a local minimum, and therefore, the structure
of Eq.(\ref{Wet}) rather than  Eq.(\ref{MN}). 

  In conclusion, Eq.(\ref{Wet}) is the minimal Langevin equation describing ST
in a general way. All the so-far reported universality classes in one-component  
synchronizing system are describable by this equation, and consequently belong 
to one of the three classes reported for it: MN1, DP or are discontinuous.
See \cite{synchro} for further details and analysis.

\section{Miscellaneous comments \label{mix}}

We present some miscellaneous remarks on issues related to the ones discussed 
along this paper; they might be of some interest in order to get a global
picture of the state of the art. See also the recent review \cite{korea}.

\begin{itemize}

\item {\bf Negative Diffusion}
 
By considering negative  values of the diffusion constant, 
anti-ferromagnetic ordering can be studied. In particular, Birner \ea
\cite{Birner} have recently studied a MN equation
with $p=3$. They consider both positive as well as negative values of the
order parameter field and report on an {\it anti-ferromagnetic} MN
phase transition (in the MN1 class).

\item {\bf Additive Noise} 

The presence of additive noise in an otherwise MN changes the 
universality of the transition \cite{NIOT,ReTr,GMS}: it is a relevant 
perturbation from the RG point of view. This is easily checked 
as the additive noise is more relevant from naive power-counting 
analysis than the multiplicative one. 
The phenomenon of {\it reentrance} may appear when both
multiplicative and additive noises act together: the system can become more ordered
by increasing the MN amplitude, but eventually the 
additive part will take over, disordering the system \cite{ReTr,GMS}. 
For example, it has been shown that  annealed Ising  models, known to have a reentrant
transitions can be described by Langevin equations with competing  additive and
multiplicative noises \cite{GMG}.

\item {\bf Colored Noise}

 One can wonder what is the effect of colored noise in MN type of 
equations. From the RG point of view deviations from Gaussianity in the noise are 
irrelevant, but they might have some effects on the phase diagram (see for
instance \cite{Wio,Bar}).
In \cite{Kramer} the effect of spatially  oscillating MN has been studied.

\item {\bf Higher order Noises}

  Having studied Langevin equations with minimal potentials and 
noises amplitudes proportional to $\phi^0$,  $\phi^{1/2}$, and   $\phi^1$, one might wonder
if higher powers, as for instance $3/2$ or $2$ have any  physical significance.
If one construct using standard Fock space techniques \cite{Fock} an effective
Langevin equation (or equivalently  a generating functional) for reaction diffusion 
process occurring in n-plets (triplets, quadruplets, etc), it is easy to see that 
they generate noise amplitudes proportional to $\phi^{n/2}$ 
If all possible reactions in a given system require the presence of triplets, 
it is reasonable to expect a $3/2$ type of noise  
\cite{nplet,nplet2}. For a general $n-plet$ process a noise amplitude
 proportional to  $\phi^{n/2}$ is expected to be the leading contribution. 
This could generate new universality classes not fully explored yet.
See also \cite{Denisov} where other noise terms are studied.

\item  {\bf Long range interaction}

If one considers long-range (for example, van der Waals) 
interactions between the substrate and
the interface in the wetting problem
the potential has the general form \cite{Lip,Rev-wet}
\begin{equation}
V(h)=b(T) h^{-m}+c h^{-n}, \qquad n>m>0.
\end{equation}
A similar interaction has been recently used to describe a class of synchronization
problems for which a new universality class has been claimed to appear \cite{LD}.
We are presently working in this problem.

\item  {\bf Quenched disorder}

 The presence of quenched disorder is known to generate new non-trivial phenomenology
in both Ising and DP -like systems. It is also expected to be a relevant perturbation
at the different MN universality classes reported here. Its analysis and relation with
physical processes, as localization, is currently under investigation.

\item {\bf Multicritical Points} 

At the points where different critical lines meet,
a different, multicritical behavior is expected to show up. 
These multicritical points and their
corresponding Langevin representations 
(most likely including MN) 
are still to be studied in depth \cite{Ahlers,LD}. 
 A multicritical point like these should be related to 
critical non-equilibrium wetting as described in \cite{Lisboa2} (see also
\cite{last}).

\item {\bf Complex Multiplicative Noise}

  It is well known that certain (annihilation) processes can be represented
by a Langevin equation with MN, but a complex amplitude
\cite{Peliti,Lee,Fock,Gardiner,nature,Tauber}. The degree of validity of 
such Langevin equations with complex noise still lacks of a satisfactory analysis; 
see \cite{nature,Tauber} for some discussions of this and related topics.
 
\end{itemize}

\section{Conclusions}

 We have reviewed different aspects of non-equilibrium phase transitions exhibited by
systems amenable to be represented by a Langevin equation with MN.
 These systems exhibit a phase transition from an absorbing phase to an active one, as
for example pinning-depinning transition, non-equilibrium wetting transitions and
synchronization transitions.  A rich phenomenology can be further explored by
exploiting the Cole-Hopf transformation that maps MN systems into 
non-equilibrium (KPZ) interfaces in the presence of an extra limiting wall. 
Depending on
the position and nature of the wall we have identified and characterized 
up to 4 different types of
critical behavior that are synthesized in Table 4.  

 It is our hope that this review will foster new developments in the fields of 
non-equilibrium phase transitions and systems with MN, and 
stimulate the application of these ideas to other open problems 
in physics and in other disciplines.

\vspace{0.5cm}
{\center \bf \Large Appendix A: Ito versus Stratonovich}
\vspace{0.5cm}

Let  us consider a MN equation in the Stratonovich representation.
Then it is completely analogous to an Ito-interpreted Langevin equation in which 
the linear coefficient $a$ has to be  replaced by $a + \sigma^2/2$ 
\cite{VK,Gardiner}. Therefore,  for this type of noise the only difference between Ito and
Stratonovich tantamounts to a shift in the control parameter. 
Along this paper we use both Ito and Stratonovich interpretations, as the results
obtained for any of them can
be straightforwardly modified to  apply to the other. 
An exception are noise induced transitions as
those reported in \cite{NIOT}, the difference between Ito-Stratonovich 
is essential in this case, as the noise amplitude plays a
key role in the linear control parameter only if the equation is interpreted a la 
Stratonovich \cite{GMS}. See \cite{Carrillo} where a new type of   
 noise induced transition is reported to occur for both Ito and Stratonovich 
interpretations.

{ {\center \bf \Large Appendix B: The 0-dimensional case. Solution of the stationary Fokker-Planck
equation. Comparison with RFT.} 
\vspace{0.5cm}

 Let us consider Eq.(\ref{MN}) in the zero-dimensional limit
\begin{equation}
{\partial}_{t} \phi(t) = - a \phi - b \phi^{p} +  \sigma  \phi \eta
\label{0d}
\end{equation}
The associated Fokker-Planck equation in the Stratonovich representation is 
\begin{equation}
\partial_t P(\phi,t) =
\partial_\phi [ (a \phi + b \phi^p) P(\phi)] + {\sigma^2 \over 2} \partial_\phi 
 [ \phi \partial_\phi [\phi P(\phi)]]
\end{equation}
whose stationary solution (trivially obtained imposing detailed balance) 
is given by Eq.(\ref{MN0d}). 
As long as $a>0$ the probability becomes non-normalizable, and the only  stationary  
state is $\phi=0$. Therefore the transition is located at $a_c=0$. 
In the Ito interpretation it is shifted to $a= \sigma^2/2$. 
Observe that in the absorbing phase the singularity  at the origin is non-integrable 
and changes continuously its degree but, above the critical point,
 the singularity disappears. 
Instead, for the RFT noise (\ie replacing $\phi$ by $\phi^{1/2}$ in the noise 
term), following the same steps one obtains,
\begin{equation} 
P_{st}(\phi)= {1 \over \phi} \exp{\left[-{2 \over \sigma^2}
 \left( a \phi - {\phi^p  \over p}
\right)\right] }.
\end{equation}
Here the singularity exists in both the active and the absorbing phase,
therefore $P(\phi)$ reduces to a delta function at the origin.
This reflects the fact that for RFT-like finite systems the only true stationary state
is the absorbing one. 
Different plots, comparisons and discussion of the (pseudo)-stationary potentials
for RFT and MN can be found in \cite{nature}.

 {\center \bf \Large Appendix C: The problem of the quenched KPZ}
\vspace{0.5cm}

 Let us consider the problem of a KPZ interface, with a {\bf positive non-linearity} 
in the presence of quenched disorder (QKPZ), $\xi(x,h)$, where $\xi$ 
is a Gaussian distribution. It was proposed in \cite{QKPZDP} that the scaling 
properties of this problem at criticality should be controlled by a DP 
type of fixed point in $d=1$
\footnote{To be more precise the QKPZ is believed to be described by 
{\it directed percolation depinning}. See \cite{Laszlo} and \cite{HZ} and references therein.}. 
The logic behind this prediction is that the interface is pinned by a 
{\it path of impurities} such that it percolates from one side to the other, blocking
the whole interface front.  
As the path is directed, DP is expected to control its scaling. Observe that,
owing to topological reasons the same reasoning does not apply to two dimensional systems. 
In $d=2$ a {\it directed surface} is required to block the interface advance. A scaling
theory of directed surfaces was proposed in \cite{surfaces} to account for this.

Despite of the numerous computational confirmations of the QKPZ-DP connection, no 
analytical study has succeeded so far in establishing such a link in a more
rigorous way. The difficulties stemming from the fact that it seems that the order 
parameter has to be changed: it is the interface velocity for KPZ, while it should be 
related to the density of pinning sites in the QKPZ case. Establishing under
firm grounds the equivalence between QKPZ and DP remains an open challenging problem.

 In the case of a {\bf negative KPZ non-linearity}, simulations show
either triangular patterns resembling the ones discussed previously 
\footnote{Named ``facet'' in the context of surface growth.}
and a first order pinning-depinning transition has been reported 
to occur \cite{Korea} or, alternatively, critical properties
controlled by DP \cite{Korea2} depending upon model details or parameter values.
Let us emphasize the deep analogies between this situation and the one described 
in Section \ref{wet}.  

{\center \bf \Large Acknowledgments}
\vspace{0.5cm}

I gratefully acknowledge my collaborators on the issues discussed in this review.
I am particularly indebted with Geoff Grinstein and Yuhai Tu, who initiated me into 
this field and taught me much about this and other issues, 
and also with
Terry Hwa, Walter Genovese, Margarida Telo da Gama, Pedro L. Garrido, 
Romualdo Pastor-Satorras, Jos\'e M. Sancho, and Abdelfattah Achahbar.
I also thank Ron Dickman, Roberto Livi, Luciano Pietronero, 
Giorgio Parisi, Matteo Marsili, Haye Hinrichsen, Hugues Chat\'e, 
Raul Toral, Pablo Hurtado, Andrea Gabrielli, Claudio Castellano,
Francesca Colaiori, and Lorenzo Giada for illuminating discussions. 
I thank Francisco de los Santos for enjoyable collaborations
as well as for a critical reading of the manuscript.
My warmest thanks also to the editors, Elka and Rodolfo, for inviting me
to participate in this book.
Financial support from the Spanish MCyT (FEDER) under project BFM2001-2841 
is also acknowledged.

\end{document}